\documentclass[11pt,twoside]{article}

\usepackage{asp2006}
\usepackage{epsf}
\usepackage{psfig}
\usepackage{lscape}
\usepackage{graphicx}
\usepackage{natbib}
\usepackage{epstopdf}

\newcommand{\msun}{M$_{\odot}$}

\newcommand{\teff}{T$_{eff}$}
\newcommand{\teq}{T$_{eq}$}

\newcommand{\meth}{CH$_4$}
\newcommand{\wat}{H$_2$O}
\newcommand{\ammon}{NH$_3$}

\newcommand{\hh}{H$_2$}
\newcommand{\nn}{N$_2$}
\newcommand{\kms}{km~s$^{-1}$}

\begin{document}
\title{The Brown Dwarf-Exoplanet Connection}   
\author{Adam J.\ Burgasser}   
\affil{Massachusetts Institute of Technology}    

\begin{abstract} 
Brown dwarfs are commonly regarded as easily-observed templates 
for exoplanet studies, with comparable masses, physical sizes and atmospheric properties. There is indeed considerable overlap in the photospheric temperatures of the coldest brown dwarfs (spectral classes L and T) and the hottest exoplanets.
However, the properties and processes associated with brown dwarf and exoplanet atmospheres can differ significantly in detail; photospheric gas pressures, elemental abundance variations, processes associated with external driving sources, and evolutionary effects are all pertinent examples.  In this contribution, I review some of the basic theoretical and empirical properties of the currently known population of brown dwarfs, and detail the similarities and differences between their visible atmospheres and those of extrasolar planets.  I conclude with some specific results from brown dwarf studies that may prove relevant in future exoplanet observations.
\end{abstract}


\section{A Brown Dwarf Primer}

Brown dwarfs are stellar objects with insufficient
mass to sustain core hydrogen fusion reactions, resulting in a steady
decline in both luminosity and effective temperature ({\teff}) with time.
The mass limit for sustained hydrogen fusion is roughly 0.072~{\msun} (75 Jupiter masses) for a Solar metallicity gas mixture, increasing to 0.090~{\msun} for a pure hydrogen gas (e.g., \citealt{2000ARA&A..38..337C}).  This mass limit establishes a formal division between ``stars'' and ``brown dwarfs'', although   such a division is not necessarily relevant to how these objects form. 
While there is ongoing debate over the details of brown dwarf formation
(the roles of gas turbulence, fragmentation and dynamical interactions; see recent reviews by \citealt{2007prpl.conf..443L} and \citealt{2007prpl.conf..459W}),
observational evidence indicates brown dwarfs are created in a manner
similar to, or at least coincident with, stars, via gravitational
collapse of dense cores within giant molecular clouds.
As a brown dwarf's energy reservoir arises primarily from the gravitational potential energy released in their initial contraction,\footnote{Small contributions also arise from brief periods of lithium- and deuterium fusion for objects more massive than $\sim$0.065~{\msun} and $\sim$0.012~{\msun}, respectively.  The latter limit is considered a possible dividing line between ``brown dwarfs'' and ``planets'' (see \citealt{2006AREPS..34..193B}), an issue that will not be touched upon here.} the luminosity, {\teff} and emergent spectral energy distribution of a brown dwarf depend primarily on mass and age, and secondarily on elemental abundances, bulk properties (e.g., rotation) and external drivers (e.g., the presence of close companion).  The interdependence of these factors on brown dwarf observables challenges the characterization of individual sources in the well-mixed Galactic population; however, it also provides an opportunity to study 
a broad range of low-temperature atmospheric properties and processes.

\begin{figure}[hbt]
\begin{center}
\includegraphics[width=3.5in,angle=90]{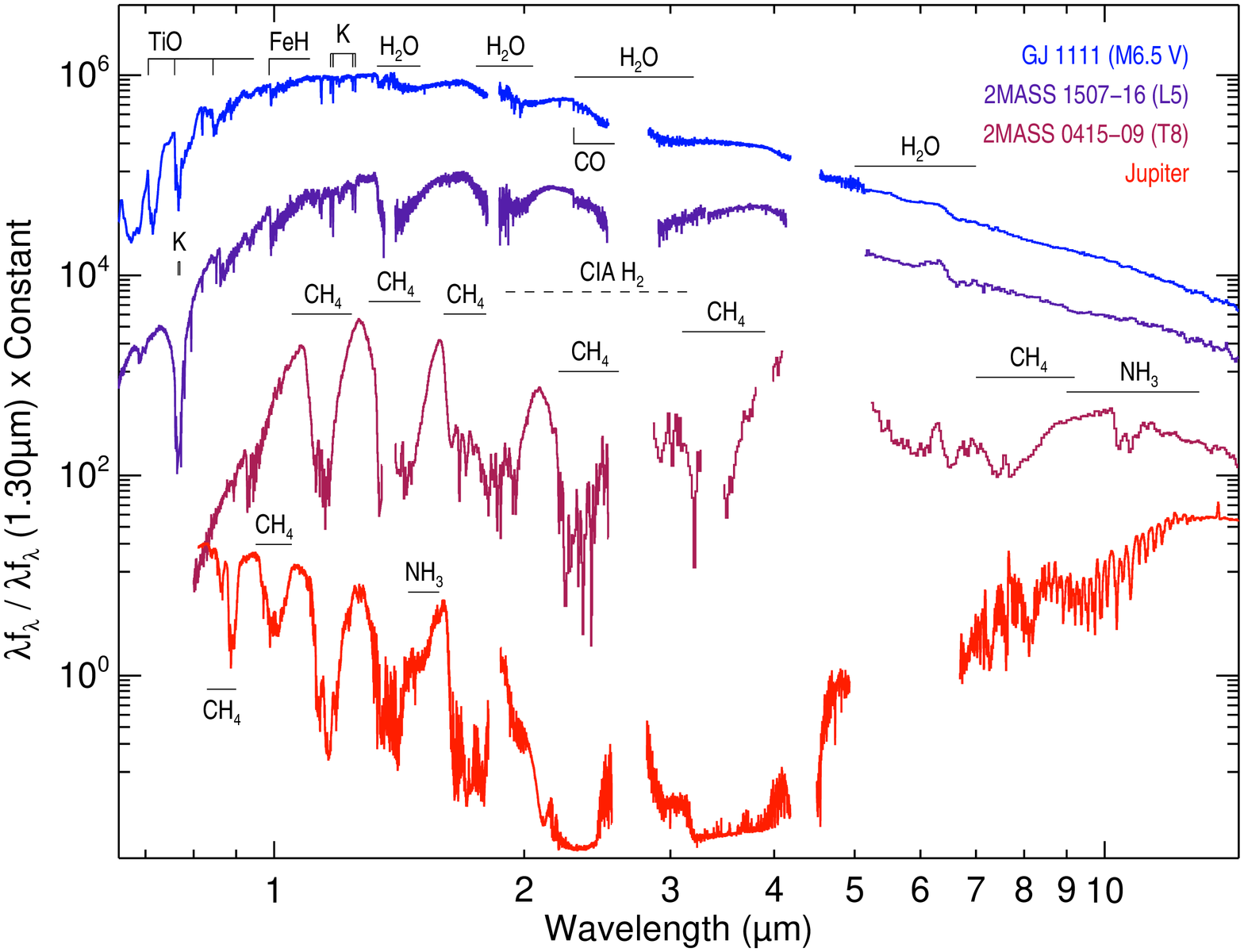}
\caption{Observed optical to mid-infrared (0.65--14.5~$\micron$) spectra of representative M-type,
L-type, and T-type dwarfs, 
compared to data for Jupiter (top to bottom).  Dwarf spectra are from \citet{2006ApJ...648..614C} and 
references therein; Jupiter data are from \citet{rayner} and \citet{2004Sci...305.1582K}.
Spectra are arbitrarily normalized.  Major molecular absorption
bands characterizing these spectra are labeled, including TiO, FeH, {\hh}, {\wat}, CO, {\meth} and {\ammon}.  Atomic K~I absorption is also labeled, which produces a substantial pressure-broadened line feature spanning 0.7--0.85~$\micron$ in L and T dwarf spectra.  Note that Jupiter's emission  
shortward of $\sim$4~$\micron$ is dominated by scattered solar light modulated
by {\meth} and {\ammon} absorption features, while the dwarf spectra are entirely emergent flux (from \citealt{mar08}). }
\label{fig1}
\end{center}
\end{figure}

Brown dwarfs have been directly observed since the mid-1990s,\footnote{On a historical note, both the discovery of the first widely-accepted brown dwarf, Gliese 229B \citep{1995Natur.378..463N}, and the discovery of the first extrasolar gas giant planet, 51 Peg b \citep{1995Natur.378..355M},
were announced to the community in the same conference, Cool Stars 9, in October 1995; see
\citet{2000prpl.conf.1313O} for a historical review.} and there are now hundreds known to exist in young clusters, as companions to nearby stars, and, most commonly, as faint isolated systems within a few hundred parsecs of the Sun.  The currently known population is segregated into
three spectral classes based on the morphology of their optical or near-infrared spectra: M dwarfs, L dwarfs and T dwarfs (Figure~\ref{fig1}).  M dwarfs encompass the warmest, youngest, and most massive brown dwarfs which have had little time to cool.  They exhibit spectral traits similar to older, low-mass dwarf stars, with strong metal-oxide molecular bands (including TiO, VO, CO and {\wat}) and neutral atomic line absorption blanketing their emergent spectral energy distributions.  L dwarf spectra are characterized by strong metal-hydride (FeH, CrH), {\wat} and CO molecular absorption; and alkali lines, including the heavily pressure-broadened Na~I and K~I doublets that largely sculpt the optical spectra of these sources (e.g., \citealt{2003A&A...411L.473A,2003ApJ...583..985B}).  L dwarfs also show evidence of condensate clouds in their photospheres, which give rise to highly reddened spectral energy distributions and absorption features from silicate grains (\citealt{2006ApJ...648..614C}; see $\S$3.1). T dwarfs are the coldest class of brown dwarfs currently known, characterized by {\wat}, {\meth}, {\ammon} and strong collision-induced {\hh} absorption.  T dwarfs do not appear to have abundant condensate material in their photospheres.  A fourth spectral class, the Y dwarfs, has been proposed for brown dwarfs even cooler than class T, although there is as yet no consensus on the general properties of this class nor a widely-accepted prototype (see \citealt{2008A&A...482..961D,2008MNRAS.391..320B}).
The M, L and T spectral classes coincide roughly with {\teff} ranges of $\ga$ 2400~K, 2400 $\la$ {\teff} $\la$ 1400, and 1400 $\la$ {\teff} $\la$ 600, respectively \citep{2004AJ....127.3516G,2004AJ....127.2948V}, although the end-point of the T spectral class remains uncertain. Variations in secondary parameters, such as metallicity, age and cloud properties modulate this temperature scale
(e.g., \citealt{2006ApJ...639.1095B,2006ApJ...651.1166M,2008ApJ...674..451B}). For more information on the L and T spectral classes, see the recent review of \citet{2005ARA&A..43..195K}.

Molecules are a prominent feature of brown dwarf atmospheres and are fundamental in our ability to ascertain the physical properties of 
individual sources.  Beyond spectral classification,
the presence, relative strengths and detailed shapes of molecular features observed in brown dwarf spectra enable measures of {\teff}, surface gravity, metallicity, cloud composition, atmospheric dynamics, rotation, and even the presence of unseen companions
(e.g., \citealt{1999ApJ...525..466L,2006ApJ...639.1095B,2006ApJ...647..552S,2008ApJ...674..451B,2008ApJ...678.1372C,2008ApJ...684.1390R}).  
Extracting these details for individual brown dwarfs is a current topic of interest in the field, and a challenge due to persistent 
inadequacies in theoretical spectral models and opacity line lists.  The complex
opacities of warm molecular gases and strongly pressure-broadened atomic features
(e.g., \citealt{2008ApJS..174..504F}; also see contribution by Tennyson),
dynamical effects on gas chemistry (e.g., \citealt{1999ApJ...519L..85G}), and the complex
processes associated with condensate grain formation (e.g., \citealt{2001ApJ...556..872A,2006A&A...455..325H}) are major hurdles in bringing atmospheric models into detailed agreement with observational data.  Progress is
being made on the theoretical front through new work on grain 
formation (e.g. \citealt{2008ApJ...675L.105H}; see contributions by Allard and Freytag), quantum opacity calculations for key molecules (e.g., \citealt{2006MNRAS.368.1087B}), and incorporation of nonequilibrium 
chemistry
(e.g., \citealt{2006ApJ...647..552S}; see contribution by Homeier).  On the observational side, the identification of
benchmark sources---companions to age-dated stars, coeval cluster members, and resolved astrometric and eclipsing binaries---are a priority
as critical tests of advanced models (e.g., \citealt{2004ApJ...609..885M,2004ApJ...615..958Z,2008ApJ...682.1256L,2008arXiv0807.2450D}).

\section{Comparing Exoplanets to Brown Dwarfs}

The benefit of brown dwarfs to exoplanet studies lies in our current ability
to study their atmospheres in considerable detail, over a broad
range of wavelengths and spectral resolutions, and over time.  Yet for brown dwarfs to be used as reliable templates for exoplanetary studies, it is essential to first assess whether their emergent spectra faithfully guide our interpretations of emergent/reflectance planetary spectra.
To this end, I examine some of key similarities and differences in the physical properties and processes of brown dwarf and exoplanet atmospheres.

\begin{figure}[hbt]
\begin{center}
\includegraphics[width=4.7in]{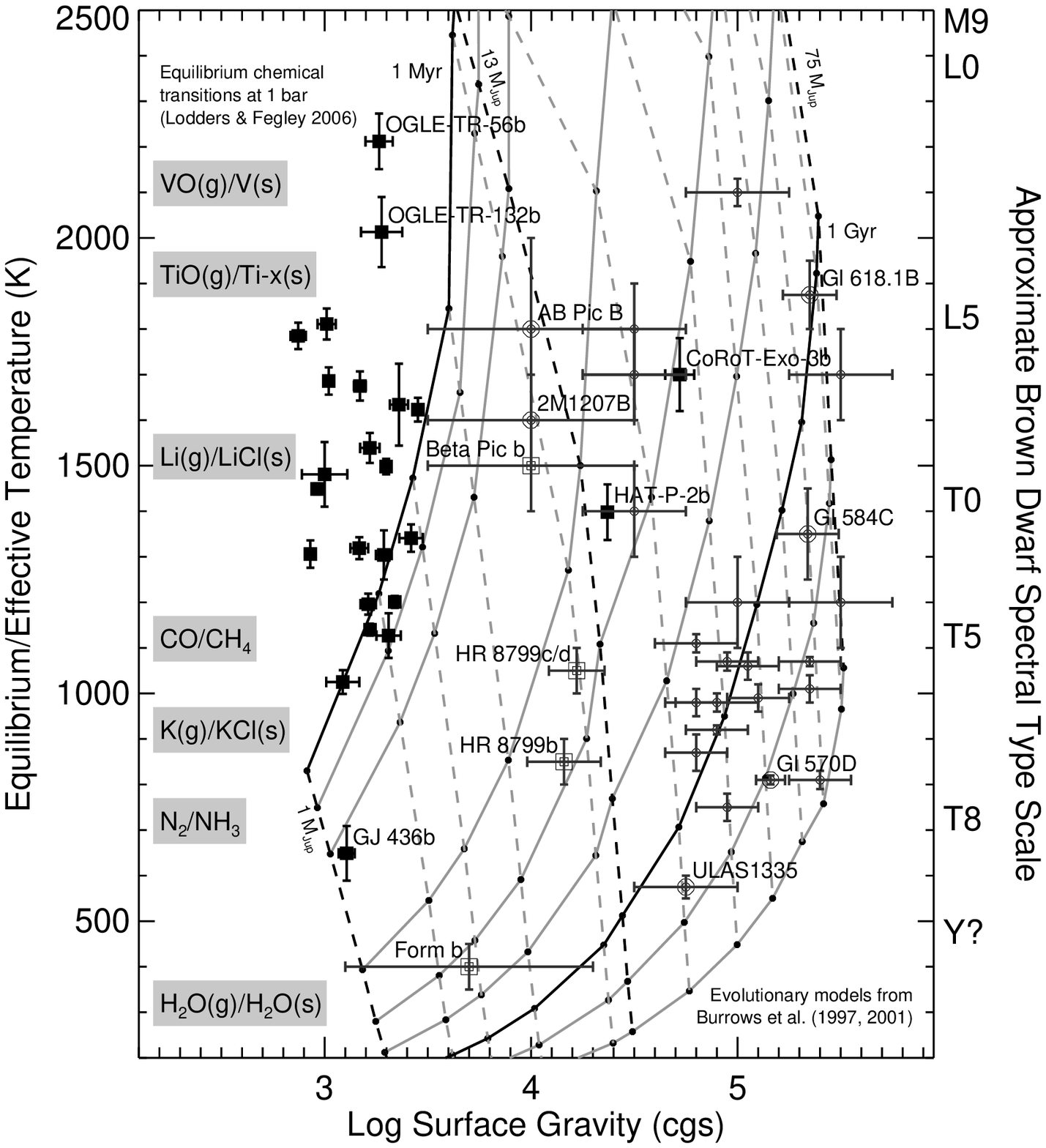}
\caption{Atmospheric gas properties of extrasolar planets and brown dwarfs,
as traced by {\teff}/{\teq} and surface gravity.  
{\teq} and $g$ values for transiting planets are from \citet{2008ApJ...677.1324T} and indicated by solid black squares (outliers are specifically labeled).
Inferred {\teff} and $g$ parameters for the directly detected planets, HR 8799bcd \citep{2008arXiv0811.2606M}, Fomalhaut b \citep{2008Sci...322.1345K}, and $\beta$ Pictoris b \citep{2008arXiv0811.3583L} are indicated by open squares
and labeled.  {\teff} and $g$ parameters for several field
brown dwarfs \citep{2006ApJ...639.1095B,2008ApJ...678.1372C} and a sample of benchmark sources (labelled; \citealt{2007ApJ...657.1064M,2001AJ....121.3235K,2001AJ....122.1989W,2006ApJ...647..552S,2005A&A...438L..29C,2008MNRAS.391..320B}) are indicated by
open circles.  These measurements are compared to the brown dwarf (``hot start'') evolutionary models of \citet{1997ApJ...491..856B,2001RvMP...73..719B}.  Solid lines delineate ages of 1, 5, 10, 30, 100, 300 Myr and 1, 3, and 10 Gyr, with 1 Myr and 1 Gyr isochrones highlighted.  Dashed lines delineate 
masses of 0.001, 0.002, 0.003, 0.005, 0.01, 0.012, 0.02, 0.03, 0.04, 0.05, 0.06, 0.07 and 0.072 {\msun}, with the 0.001 {\msun} = 1 Jupiter mass, 0.012 {\msun} = 13 Jupiter masses (deuterium burning limit) and 0.072 {\msun} = 75 Jupiter masses (hydrogen burning limit) lines highlighted.  An approximate spectral classification scale based on {\teff} determinations by \citet{2004AJ....127.2948V} is indicated on the right axis.
Equilibrium chemical transitions for several key species are
indicated along the left side of the plot \citep{2006asup.book....1L}.
 }
\label{fig2}
\end{center}
\end{figure}

\subsection{Temperatures}

A gross assessment of the photospheric temperatures of brown dwarfs can be inferred from their {\teff}s.  These are typically determined from bolometric luminosity measurements and an assumed (theoretical) radius estimate (e.g., \citealt{2004AJ....127.3516G,2004AJ....127.2948V}); alternately, fits of spectral data to theoretical models are used (e.g., \citealt{2004ApJ...609..854M,2006ApJ...639.1095B,2008ApJ...678.1372C}).  These measures do not always agree \citep{2003ApJ...599L.107S}.  For planets, a comparable statistic is the thermal equilibrium temperature, {\teq} = T$_*$(R$_*$/2$a$)$^{1/2}$, where T$_*$ and R$_*$ are the effective temperature and radius of the host star, respectively, and $a$ the semi-major axis (ignoring albedo and orbital eccentricity). As it turns out, the {\teff}s of L- and T-type brown dwarfs overlap considerably with the {\teq}s of transiting extrasolar planets (Figure~\ref{fig2}).  Similarly, the directly-imaged planets HR~8799bcd \citep{2008arXiv0811.2606M}, Fomalhaut b \citep{2008Sci...322.1345K} and $\beta$ Pictoris b \citep{2008arXiv0811.3583L} have estimated {\teff}s (not {\teq}s; see below) comparable to T dwarfs.  Fomalhaut b may in fact be cooler than the {\teff} = 575$\pm$25~K ULAS~1335, the coldest brown dwarf currently known \citep{2008MNRAS.391..320B}.  

Transiting planets are warm
due to the radiative forcing by their host stars.  A planet with {\teq} = 500~K lies only 0.3~AU (0.07~AU) from a solar-type (M0 dwarf) primary.  As {\teq} $\propto$ T$_*$a$^{-1/2}$, more widely-orbiting planets and planets orbiting less luminous host stars have lower {\teq}s, below the range currently sampled by brown dwarfs.  For closely-orbiting, tidally-locked hot Jupiter planets, care must be taken when using {\teq}
as a proxy for photospheric temperature, as these planets can have substantial day/night asymmetries (see contribution by Knutson).  Eccentricity effects can also give rise to large temporal modulations in {\teq} (see contributions by Iro and Lewis).  In contrast, HR 8799bcd, Fomalhaut b, and $\beta$ Pictoris b have (to first order) uniformly warm photospheres dominated by internal heat rather than reprocessed host star light.  These planets are still young ($<$300~Myr); like
brown dwarfs, their atmospheres will eventually cool to low temperatures.

\subsection{Photospheric Pressures}

While the mean photospheric gas temperatures of brown dwarfs and planets are comparable, gas pressures are generally quite different.  At the photosphere, gas pressure is proportional to the surface gravity, $g$, as $P_{ph} \propto g/\kappa$, where $\kappa$ is the Rosseland mean opacity.  Evolutionary models dictate that the surface gravities of brown dwarfs depend strongly on mass (due to their nearly constant radii) and weakly on age (significant variations only for ages $\la$ 100~Myr; see Figure~\ref{fig2}).  Surface gravities for evolved  M, L and T dwarfs (ages $\sim$0.5--10~Gyr) span g$\sim$300--3000~m~s$^{-2}$.  In contrast, the vast majority of transiting exoplanets have g$\sim$10--30~m~s$^{-2}$, as directly inferred from radial velocity and transit light curves (e.g., \citealt{2007ApJ...664.1190S}). Ignoring opacity effects, the photospheric gas pressures of transiting exoplanets are 1--2 orders of magnitude less than those of brown dwarfs.  Cooler, widely-orbiting Jupiter-mass gas giants also have photospheric pressures about 10 times less than their (typically more massive) brown dwarf counterparts. 

Differences in photospheric gas pressure can have a measurable influence on some chemical pathways.  One example is the reduction reaction CO + 3{\hh} $\rightarrow$ {\meth} + {\wat}, which favors {\meth} production in high-pressure gas environments.  Chemical equilibrium models indicate that {\meth} becomes abundant in brown dwarf photospheres (1-10 bar) below $\sim$1400~K, but in planetary photospheres (0.01--1 bar) below $\sim$900~K (e.g., \citealt{2006asup.book....1L}).  Pressure also modulates some gas opacities, notably the pressure-broadened alkali lines that dominate the optical spectra of L and T dwarfs, and collision-induced {\hh} absorption that suppresses broad swaths of infrared light in the coldest brown dwarfs (e.g., \citealt{1969ApJ...156..989L,1994ApJ...424..333S}).  Both features are used to constrain surface gravities for individual brown dwarfs near the Sun (e.g., \citealt{1999AJ....118.2466M,2006ApJ...639.1095B,2006ApJ...639.1120K}) and verify the membership of brown dwarfs in young clusters (e.g., \citealt{1999ApJ...525..466L,2007ApJ...657..511A}). 

Fortuitously, there is overlap in photospheric gas pressure/temperature space between the youngest and lowest-mass brown dwarfs---those found in young star-forming regions and associations---and dense gas giant planets that are either very massive or have a substantial core.  Transiting planets such as HD 147506b (aka Hat-P-2b; $\rho \approx$ 13~g~cm$^{-3}$; \citealt{2007ApJ...670..826B}) and CoRoT-Exo-3b ($\rho \approx$ 26~g~cm$^{-3}$; \citealt{2008A&A...491..889D})
have surface gravities similar to 30--100~Myr, $\sim$5--20 Jupiter mass brown dwarfs like 2MASS 1207-39B and AB Pic B (\citealt{2004A&A...425L..29C,2005A&A...438L..29C,2007ApJ...657.1064M}; see Figure~\ref{fig2}).
With a mass of 22~Jupiter masses, CoRoT-Exo-3b could be properly classified as a highly irradiated brown dwarf companion.  

In addition to mean values of photospheric temperature and pressure, differences  in the
pressure-temperature profiles of brown dwarfs and exoplanets must be considered.  For planets, external heating from the host star flattens out the pressure-temperature profile and can give rise to inversion layers.  This  translates into  variations in the local gas chemistry and changes in the atmospheric column abundances of atomic and molecular absorbers.
In addition, strongly irradiated planets develop deep radiative envelopes that extend well below the visible photosphere, whereas brown dwarf atmospheres are fully convective through to their photospheres (e.g., \citealt{1997ApJ...491..856B}). Differences in the gas mixing rates and vertical temperature profiles between externally heated planetary atmospheres and brown dwarf atmospheres can produce profound differences in emergent spectral energy distributions, even for sources with comparable photospheric gas temperatures and pressures (e.g., \citealt{2008ApJ...678.1419F}).

\subsection{Compositions}

The elemental composition, or metallicity, of a cool atmosphere also modulates chemistry and spectral appearance.   Gas-giant planets tend to have metal-rich atmospheres, having condensed out of the gas-depleted debris disks around preferentially metal-enriched host stars (e.g., \citealt{1997MNRAS.285..403G}).  Ice-giant (i.e., Neptune) and terrestrial planet atmospheres exhibit even greater metallicity enhancements.  These trends are present in the solar system: the atmospheres of Jupiter, Saturn, Uranus and Neptune have effective metal abundances ranging from $\sim$3 to $\sim$40 times that of the Sun \citep{2007Ap&SS.307..279F}.  More importantly, there is a large range in individual elemental abundances, driven by the segregation of volatiles in the Sun's early protoplanetary disk and chemical separation in planetary atmospheres (e.g., He settling in Saturn).  Significant variations in elemental abundances can have as great or greater impact on the chemistry and molecular composition of cool atmospheres as pressure or temperature alone (e.g., \citealt{2007ApJ...654L..99T,2008ApJ...683.1104F}).  

In contrast, brown dwarf metallicites are expected to span the same range
as stars, topping out at perhaps 3--5 times solar abundances
but extending down to significantly subsolar abundances in the metal-poor thick disk and halo populations.  For example, members of the recently-identified L subdwarf class have metallicities $\sim$0.01-0.1 times solar \citep{2007ApJ...657..494B,2008arXiv0811.4136S}.  Even brown dwarfs with bulk solar metallicities will have slightly metal-poor photospheres due to condensation effects.  Like stars, relative elemental abundances of brown dwarfs are likely to exhibit only small variations, although condensation effects may modify abundance patterns. In any case, the broad range of elemental abundances observed in the solar planets and expected to a greater degree among the wider exoplanet population will probably not be realized among Galactic brown dwarfs.

\subsection{Stellar Hosts and Driving Forces}

Unlike the majority of brown dwarfs, exoplanets are generally accompanied by a luminous host star, which ultimately maintains its atmosphere in a warm state, modulo variations arising from orbital eccentricity, circulation, or magnetic interaction effects.  Radiation and stellar winds drive non-equilibrium dynamics in exoplanet atmospheres, including internal winds/jets (see contribution by Showman) and atmospheric stripping (see contribution by Alyward).  UV and X-ray radiation drive photochemical production of hazes in exoplanet atmospheres (see contribution by Yung), and the formation of upper inversion layers.  Tidal locking from a close stellar companion slows an exoplanet's rotation and can provide a source of (temporary) internal heating.  

These processes will not generally occur in brown dwarf atmospheres.
The hottest and youngest brown dwarfs do exhibit high-energy nonthermal emission (X-ray and UV) arising from magnetic activity or accretion.  However, with the exception of rare massive flares in which the total magnetic energy output can briefly exceed thermal emission (e.g., \citealt{1999ApJ...519..345L}), high-energy magnetic emission is typically a small fraction ($<$10$^{-3}$) of the total energy budget and does not significantly alter pressure-temperature profiles. Furthermore, for cooler L and T-type brown dwarfs, magnetic emission is conspicuously absent due to the loss of field/gas coupling in the highly neutral photospheres of these objects \citep{2002ApJ...571..469M,2002ApJ...577..433G}.  Only the very youngest ($<$1~Myr), actively accreting brown dwarfs and sources in close orbits around luminous companions are likely to show significant modification of their pressure-temperature profiles as a result of external driving forces.

The absence of both radiative and mechanical forcing on brown dwarf atmospheres is particularly relevant to atmospheric circulation and dynamics (see contributions by Showman and Cho). Rotational modulation of weather phenomena has been invoked to explain low-level spectral and photometric variability detected in some brown dwarfs (see review by \citealt{2005AN....326.1059G}).  The timescales for these variations are generally consistent with rotational line broadening measurements (e.g., \citealt{2008ApJ...684.1390R}) and variations in nonthermal magnetic emission (e.g., \citealt{2006ApJ...648..629B}). The general absence of magnetic field coupling (i.e., spots) and the clear presence of condensate clouds makes weather an appealing explanation for this variability, particularly given observed long-term period variations and changes in variability amplitudes.  However, the
winds and jets that drive weather in planetary atmospheres arise from asymmetric radiative forcing by the host star; such forces are absent for most brown dwarfs.  It is possible that winds could be driven by the extremely rapid rotations of brown dwarfs.  Periods of 1--10~hours are typical (cf.\ Jupiter's 11-hour period) and surface rotational velocities of up to 80~{\kms} have been measured \citep{2008ApJ...684.1390R}.  This is an order of magnitude faster than the rotations of tidally-locked giant planets, and as a result coriolis forcing is more important in brown dwarf atmospheres.  However, because of their greater surface gravities and photospheric pressures, the Rhines length and Rossby deformation radius scales (see contribution by Showman) in the upper atmospheres of brown dwarfs are roughly equivalent to those for hot Jupiters, of order the planetary/brown dwarf radius (assuming horizontal wind speeds comparable to the local sound speed, $\sim$1~{\kms}; see \citealt{2008ASPC..398..419S}).  As such, the small-scale banding and storm vorticities that characterize Jupiter's visible atmosphere are probably not common on either brown dwarfs or hot Jupiters, although detailed modeling of the former have yet to be reported.

\smallskip
\smallskip

In summary, while the current populations of brown dwarfs and (warm/hot)
exoplanets may have photospheres with similar 
temperatures, significant differences in gas pressures and compositions may drive markedly dissimilar molecular chemistry.  There is some overlap in temperature/pressure space between the densest/most massive exoplanets and the youngest/least-massive brown dwarfs, where meaningful comparisons in atmospheric properties and processes may be fruitfully made.  The external forcing of a host star also results in exoplanetary atmospheric processes not seen in brown dwarfs: modified pressure-temperature profiles, inversion layers, photochemical production and thermal asymmetries that drive winds and jets.  Yet at least in terms of flow dynamics, the atmospheres of both tidally-locked hot Jupiters and brown dwarfs should be weakly banded and have few small-scale vorticities in contrast to Jupiter.

\section{Detailed Brown Dwarf Results Relevant to Exoplanet Studies}

I conclude my contribution with two examples of low-temperature atmospheric processes studied in detail in brown dwarf studies but not yet sufficiently constrained in exoplanetary studies: condensate cloud formation and nonequilibrium chemistry.

\subsection{Condensate Cloud Formation}

Condensed species present in the photospheres of L dwarfs
arise naturally from equilibrium chemistry, proceeding from the more refractory species such as 
mineral oxides and silicates (below 2500~K),
to ionic salts and sulfides (below 1000~K), 
to ``organic'' condensates including
as {\wat}[s] and {\ammon}[s] (below 300~K; see review by \citealt{2006asup.book....1L}).  The presence of condensed species in L dwarfs has been 
inferred indirectly by their very red colors and muted molecular absorption
features (e.g., \citealt{2001ApJ...556..357A}).  Direct detection of silicate
grain absorption has recently been made possible by the {\em Spitzer Space Telescope}
\citep{2006ApJ...648..614C}.

That these species reside in cloud structures in brown dwarf atmospheres
has be inferred from other indicators: elemental depletion at high altitudes and the absence of condensates in T dwarf spectra.  In the first case, the gravitational settling of condensed grains removes these species from the ambient gas, preventing further chemical reactions at higher (cooler) altitudes.
For example, K~I absorption is particularly strong in T dwarf spectra, despite
the fact that K should have condensed out into silicate grains such as KAlSi$_3$O$_8$[s] (othroclase).  This reaction is inhibited by the depletion of 
Al and Si at deeper layers 
through the formation of, e.g., CaTiO$_3$[s]
(perovskite) and Al$_2$O$_3$[s] (corundum; \citealt{1999ApJ...512..843B,2006asup.book....1L}); hence elemental K persists.\footnote{Condensate depletion is also seen in the atmospheres of Jupiter and Saturn as traced by the presence of GeH$_4$[s] (germane) over SiH$_4$[s] (silane), despite the much greater elemental abundance of Si in a solar gas mixture \citep{1994Icar..110..117F}.} The absence of condensate cloud absorption in T dwarf spectra can be explained if condensates are vertically confined in cloud structures which ultimately sink below the visible photosphere (e.g., \citealt{2001ApJ...556..872A,2002ApJ...575..264T,2003ApJ...586.1320C,2004A&A...414..335W}).  As it turns out, the disappearance of condensate clouds at the transition between L and T dwarfs is quite abrupt, suggesting dynamic effects may be critical for cloud evolution (e.g., \citealt{2002ApJ...571L.151B,2006ApJ...647.1393L,2007ApJ...659..655B,2008ApJ...678.1372C}). 

The $>$500 L dwarfs observed to date reveal substantial diversity in cloud-sensitive features, including near-infrared colors ($>$1~mag scatter in $J-K$ for a given spectral subtype) and the strength of silicate grain absorption (e.g., \citealt{2000AJ....120..447K,2003ApJ...596..561M,2008ApJ...674..451B}).
These variations have been simplistically interrupted as a range in cloud ``thicknesses'' (e.g. \citealt{2008ApJ...678.1372C}), although it is likely that other properties, such as grain size distribution, grain compositions (e.g., \citealt{2008ApJ...675L.105H}) and cloud surface coverage also contribute.  The properties of
brown dwarf clouds are almost certainly tied to (interrelated) secondary parameters of age, surface gravity, metallicity, and rotation, as suggested by empirical trends (e.g., \citealt{2009AJ....137....1F}).  However, current 
atmospheric models generally treat cloud properties as an independent
model parameter, so source-to-source variations and temporal evolution can only be treated in a somewhat ad-hoc manner.
Nevertheless, there has been substantial improvement in the fidelty and complexity of condensate cloud models to address the wealth of observational data, progress that can be ported to the still underconstrained
problem of condensate clouds in hot exoplanetary atmospheres.

\subsection{Nonequilibrium Chemistry and Atmospheric Dynamics}

The contribution of Marley touches upon nonequilibrium chemistry in brown dwarf atmospheres in considerable detail, so I present only the major results here for completeness.  Nonequilibrium chemistry refers to the nonequilibrium abundances of species that occur when the timescale for diffusive gas flow is shorter than the timescales governing the relevant chemical reactions.  In brown dwarfs, the two reactions that are most affected by nonequilibrium chemistry convert CO $\rightarrow$ {\meth} and {\nn} $\rightarrow$ {\ammon}.  CO and {\nn} have strong bonds and long chemical timescales at low temperatures, so they can appear in excess, and {\meth} and {\ammon} in depletion, as a result of diffusive flows.  Such abundance anomalies are indeed observed \citep{1997ApJ...489L..87N,1998ApJ...502..932O,2006ApJ...647..552S}, and indicate diffusivity constants of 1-100 m$^2$~s$^{-1}$, in excess of flows expected from convective instabilities \citep{sau07}.  It is likely that nonequilbrium chemistry is present in exoplanetary atmospheres as well, potentially giving rise to azimuthal abundance variations in sources with large day/night asymmetries, and hence modulation of phase-resolved spectroscopy in variance with equilibrium chemistry models.  Such effects should be specifically sought for in future phase-resolved, direct spectroscopic studies of transiting exoplanets.


\acknowledgements 
I thank A.\ Showman for insightful conversations on planetary atmospheric flow scales, K.\ Lodders and M.\ Marley for consultation on theoretical topics, and M.\ Cushing for providing Figure~1. 



\end{document}